\documentclass[12pt]{article}
\begin{document}
\title{THE MACHIAN UNIVERSE}
\author{B.G. Sidharth\\
International Institute for Applicable Mathematics \& Information Sciences\\
Hyderabad (India) \& Udine (Italy)\\
B.M. Birla Science Centre, Adarsh Nagar, Hyderabad - 500 063 (India)}
\date{}
\maketitle
\begin{abstract}
We give arguments from the point of view of Gravitation as well as Electromagnetism which indicate a Machian view for the universe.
\end{abstract}
\section{Introduction}
The dependence of the mass of a particle on the rest of the universe was argued by Mach in the nineteenth century itself in what is now famous as Mach's Principle \cite{narlikar,counter}. The Principle is counterintuitive in that we tend to consider the mass which represents the quantity of matter in a particle to be an intrinsic property of the particle. But the following statement of Mach's Principle shows it to be otherwise thus going counter to ideas of locality and causality.\\
If there were no other particles in the universe, then the force acting on the particle $P$ would vanish and so we would have by Newton's second law
\begin{equation}
ma = 0\label{ea}
\end{equation}
Can we conclude that the acceleration $a$ of the particle vanishes? Not if we do not postulate the existence of an absolute background frame in space. In the absence of such a Newtonian absolute space frame, the acceleration $a$ would in fact be arbitrary, because we could measure this acceleration with respect to arbitrary frames. Then (\ref{ea}) implies that $m = 0$. That is, in the absence of any other matter in the universe, the mass of a material particle would vanish. From this point of view the mass of a particle depends on the rest of the material content of the universe.\\
Though Einstein was an admirer of Mach's ideas, his Special Theory of Relativity went counter to them. He subscribed to the concept of locality according to which information about a part of the universe can be obtained by dealing with that part alone and without taking into consideration the rest of the universe at the same time. In his words, \cite{schrodinger} "But one supposition we should, in my opinion absolutely hold fast: the real factual situation of the system $S_2$ is independent of what is done with the system $S_1$ which is spatially separated from the former." Further, causality is another cornerstone in Einstein's Physics.
\section{The Feynman Wheeler Perfect Absorber Theory}
Feynman and Wheeler \cite{fw} gave in 1945 a consistent Electrodynamics which was called the.a Perfect Absorber Theory. Later this theory was revived and reviewed by some scholars, notably F. Rohrlick \cite{rohr} and more recently by Hoyle and Narlikar in the context of the Instantaneous Action at a Distance Cosmology \cite{hoyle,nar}. Even more recently scholars like Chubykalo, Smirnov-Rueda and others have argued in favour of the IAAD Electrodynamics \cite{chub1,chub2}. However by and large these developments have been overlooked due to the success of Quantum Electrodynamics.\\
We will now argue that the Perfect Absorber Theory of Electrodynamics is meaningful in the above context and in the context of very recent work that points to a small photon mass, which nevertheless is perfectly compatible with the experimental limits \cite{bhtd,mp}. We will then see the limit in which we return to the usual Quantum Electrodynamics.\\
We begin with Classical Electrodynamics. From a classical point of view a charge that is accelerating radiates energy which dampens its motion. This is given by the well known Maxwell-Lorentz equation, which in units $c = 1$, and $\tau$ being the proper time, while $\imath = 1,2,3,4,$ is (Cf.\cite{nar}), 
\begin{equation}
m\frac{d^2x^\imath}{d\tau^2} = eF^{\imath}_{k} \frac{dx^k}{d\tau} + \frac{4e}{3} g_{lk} \left(\frac{d^3x^\imath}{d\tau^3} \frac{dx^l}{d\tau} - \frac{d^3x^l}{d\tau^3} \frac{dx^\imath}{d\tau}\right) \frac{dx^k}{d\tau},\label{e1}
\end{equation}
The first term on the right side is the usual external field while the second term is the damping field which is added ad hoc by the requirement of the energy loss due to radiation. In 1938 Dirac introduced instead
\begin{equation}
m\frac{d^2x^\imath}{d\tau^2} = e\{F^\imath_k + R^\imath_k\} \frac{dx^k}{d\tau}\label{e2}
\end{equation}
where
\begin{equation}
R^\imath_k \equiv \frac{1}{2} \{F^{\imath}_{k(ret)} - F^{\imath}_{k(adv)}\}\label{e3}
\end{equation}
In (\ref{e3}), $F_{(ret)}$ denotes the retarded field and $F_{(adv)}$ the advanced field. While the former is the causal field where the influence of a charge at $A$ is felt by a charge at $B$ at a distance $r$ after a time $t = \frac{r}{c}$, the latter is the advanced field which acts on $A$ from a future time. In effect what Dirac showed was that the radiation damping term in (\ref{e1}) or (\ref{e2}) is given by (\ref{e3}) in which an antisymmetric difference of the advanced and retarded fields is taken. Let us elaborate a little further.\\
The Maxwell wave equation has two independent solutions, one having support on the future light cone, this is the retarded solution and the other having support on the past light cone which has been called the advanced solution. The retarded solution is selected to describe the physical situation in conventional theory. This retarded solution is physically meaningful, as it describes electromagnetic radiation which travels outward from a given charge and reaches another point at a later instant. It has also been called for this reason the causal solution. On the grounds of this causality, the advanced solution has been rejected, except in a few formulations like those of Dirac or Feynman and Wheeler. In the F-W formulation, the rest of the charges in the universe react back on the original electron through their advanced waves, which arrive at the given charge at the same time as the given charge radiates its electromagnetic waves. More specifically, when an electron is accelerated at the instant $t$, it interacts with the other charges at a later time $t' = t + r/c$ where $r$ is the distance of the other charge--these are the retarded interactions. However the other charges react back on the original electron through their advanced waves, which will arrive at the time $t' - r/c = t$. In this formulation, from this point of view, there is no contradiction with causality.\\
It must be mentioned that Dirac's prescription lead to the so called runaway solutions, with the electron acquiring larger and larger velocities in the absence of an external force \cite{hoyle}. This he related to the infinite self energy of the point electron.\\
As far as the breakdown of causality is concerned, this takes place within a period $\sim \tau$, the Compton time as we will briefly see below \cite{rohr,hoyle}. It was at this stage that Wheeler and Feynman reformulated the above action at a distance formalism in terms of what has been called their Absorber Theory. In their formulation, the field that a charge would experience because of its action at a distance on the other charges of the universe, which in turn would act back on the original charge is given by
\begin{equation}
Re = \frac{2e^2 d}{3dt} (\ddot{x})\label{e4}
\end{equation}
The interesting point is that instead of considering the above force in (\ref{e4}) at the charge $e$, if we consider the response at an arbitrary point in its neighborhood as was shown by Feynman and Wheeler (Cf.ref.\cite{fw}) and, in fact a neighborhood at the Compton scale, as was argued recently by the author \cite{iaad}, the field would be precisely the Dirac field given in (\ref{e2}) and (\ref{e3}). To see this in detail, we observe that the well known Lorentz Dirac equation (Cf.\cite{rohr}), can be written as
\begin{equation}
ma^\mu (\tau) = \int^{\infty}_{0} K^\mu (\tau + \alpha \tau_0) e^{-\alpha} d\alpha\label{e5}
\end{equation}
where $a^\mu$ is the accelerator and
$$K^\mu (\tau) = F^\mu_{in} + F^\mu_{ext} - \frac{1}{c^2} Rv^\mu ,$$
\begin{equation}
\tau_0 \equiv \frac{2}{3} \frac{e^2}{mc^3} \sim 10^{-23}sec\label{e6}
\end{equation}
and
$$\alpha = \frac{\tau' - \tau}{\tau_0} ,$$
where $\tau$ denotes the time and $R$ is the total radiation rate. Incidentally this is a demonstration of the non locality in Compton time, referred to above.\\
It can be seen that equation (\ref{e5}) differs from the usual equation of Newtonian Mechanics, in that it is non local in time. That is, the acceleration $a^\mu (\tau)$ depends on the force not only at time $\tau$, but at subsequent times also. Let us now try to characterize this non locality. We observe that $\tau_0$  given by equation (\ref{e6}) is the Compton time $\sim 10^{-23}secs$. So equation (\ref{e5}) can be approximated by
\begin{equation}
ma^\mu (\tau) = K^\mu (\tau + \xi\tau_0) \approx K^\mu (\tau)\label{e7}
\end{equation}
Thus as can be seen from (\ref{e7}), the Lorentz-Dirac equation differs from the usual local theory by a term of the order of 
\begin{equation}
\frac{2}{3} \frac{e^2}{c^3} \dot{a}^\mu\label{e8}
\end{equation}
the so called Schott term. It is well known that the time component of the Schott term (\ref{e8}) is given by (Cf.ref.\cite{rohr})
$$-\frac{dE}{dt} \approx R \approx \frac{2}{3} \frac{e^2 c}{r^2} \left(\frac{E}{mc^2}\right)^4,$$
where $E$ is the energy of the particle.  Whence integrating over the period of non locality $\sim \tau_0$ the Compton time, we can immediately deduce that $r$ the scale of spatial non locality is given by
$$r \sim c\tau_0,$$
which is of the order of the Compton wavelength.\\
The net force emanating from the charge is thus given by
\begin{equation}
F^{ret} = \frac{1}{2} \left\{F^{ret} + F^{adv}\right\} + \frac{1}{2} \left\{F^{ret} - F^{adv}\right\}\label{e9}
\end{equation}
which is the acceptable causal retarded field. The causal field now consists of the time symmetric field of the charge together with the Dirac field, that is the second term in (\ref{e9}), which represents the response of the rest of the charges. Interestingly in this formulation we have used a time symmetric field, viz., the first term of (\ref{e9}) to recover the retarded field with the correct arrow of time. Feynman and Wheeler stressed that the universe has to be a perfect absorber or to put it simply, every charged particle in the universe should respond back to the action on it by the given charge.\\
There are two important inputs which we can see in the above more recent formulation. The first is the action of the rest of the universe at a given charge and the other is minimum spacetime intervals which are of the order of the Compton scale. The minimum space time interval removes, firstly the advanced field effects which take place within the Compton time and secondly the infinite self energy of the point electron disappears due to the Compton scale.\\
The Compton scale comes as a Quantum Mechanical effect, within which we have zitterbewegung effects and a breakdown of Causal Physics \cite{diracpqm}. Indeed Dirac had noted this aspect in connection with two difficulties with his electron equation. Firstly the speed of the electron turns out to be the velocity of light. Secondly the position coordinates become complex or non Hermitian. His explanation was that in Quantum Theory we cannot go down to arbitrarily small space time intervals, for the Heisenberg Uncertainty Principle would then imply arbitrarily large momenta and energies. So Quantum Mechanical measurements are an average over intervals of the order of the Compton scale. Once this is done, we recover meaningful physics. All this has been studied afresh by the author more recently, in the context of a non differentiable space time and noncommutative geometry.\\
Weinberg also notices the non physical aspect of the Compton scale \cite{weinberg}. Starting with the usual light cone of Special Relativity and the inversion of the time order of events, he goes on to add, and we quote at a little length and comment upon it, ``Although the relativity of temporal order raises no problems for classical physics, it plays a profound role in quantum theories. The uncertainty principle tells us that when we specify that a particle is at position $x_1$ at time $t_1$, we cannot also define its velocity precisely. In consequence there is a certain chance of a particle getting from $x_1$ to $x_2$ even if $x_1 - x_2$ is spacelike, that is, $| x_1 - x_2 | > |x_1^0 - x_2^0|$. To be more precise, the probability of a particle reaching $x_2$ if it starts at $x_1$ is non-negligible as long as
$$(x_1 - x_2)^2 - (x_1^0 - x_2^0)^2 \leq \frac{\hbar^2}{m^2}$$
where $\hbar$ is Planck's constant (divided by $2\pi$) and $m$ is the particle mass. (Such space-time intervals are very small even for elementary particle masses; for instance, if $m$ is the mass of a proton then $\hbar /m = 2 \times 10^{-14}cm$ or in time units $6 \times 10^{-25}sec$. Recall that in our units $1 sec = 3 \times 10^{10}cm$.) We are thus faced again with our paradox; if one observer sees a particle emitted at $x_1$, and absorbed at $x_2$, and if $(x_1 - x_2)^2 - (x_1^0 - x_2^0)^2$ is positive (but less than or $=\hbar^2 /m^2)$, then a second observer may see the particle absorbed at $x_2$ at a time $t_2$ before the time $t_1$ it is emitted at $x_1$.\\
``There is only one known way out of this paradox. The second observer must see a particle emitted at $x_2$ and absorbed at $x_1$. But in general the particle seen by the second observer will then necessarily be different from that seen by the first. For instance, if the first observer sees a proton turn into a neutron and a positive pi-meson at $x_1$ and then sees the pi-meson and some other neutron turn into a proton at $x_2$, then the second observer must see the neutron at $x_2$ turn into a proton and a particle of negative charge, which is then absorbed by a proton at $x_1$ that turns into a neutron. Since mass is a Lorentz invariant, the mass of the negative particle seen by the second observer will be equal to that of the positive pi-meson seen by the first observer. There is such a particle, called a negative pi-meson, and it does indeed have the same mass as the positive pi-meson. This reasoning leads us to the conclusion that for every type of charged particle there is an oppositely charged particle of equal mass, called its antiparticle. Note that this conclusion does not obtain in non-relativistic quantum mechanics or in relativistic classical mechanics; it is only in relativistic quantum mechanics that antiparticles are a necessity. And it is the existence of antiparticles that leads to the characteristic feature of relativistic quantum dynamics, that given enough energy we can create arbitrary numbers of particles and their antiparticles.''\\
We note however that there is a nuance here which distinguishes Weinberg's explanation from that of Dirac. In Weinberg's analysis, one observer sees only protons at $x_1$ and $x_2$, whereas the other observer sees only neutrons at $x_1$ and $x_2$ while in between, the first observer sees a positively charged pion and the second observer a negatively charged pion. Weinberg's explanation is in the spirit of the Feynman-Stuckleberg diagrams. One particle leaves $x_1$ and then travels causally to $x_2$, where $x_1$ and $x_2$ are within the Compton wavelength of the particle. But for another observer, a particle first leaves $x_2$ and travels backward in time to $x_2$.\\
Let us consider the above in the context of a non zero photon mass. Such a mass $\sim 10^{-65}gms$ was recently deduced by the author, and it is not only consistent with experimental restrictions, but also predicts a new effect viz., a residual cosmic radiation $\sim 10^{-33}eV$, which in fact has been observed \cite{bhtd,mp,newphys,mer,phys}. Such a photon would have a Compton length $\sim 10^{28}cms$, that is the radius of the universe itself.\\
This would then lead to the following scenario: An observer would see a photon leaving a particle $A$ and then reaching another particle $B$, while a different observer would see exactly the opposite for the same event - that is a photon leaves $B$ and travels backward in time to $A$, as in the Weinberg interpretation. This latter gives the advanced potential. The distinction between the advanced and retarded potentials of the old electromagnetic theory thus gets mixed up and we have to consider both the advanced and retarded potentials. We consider this in a little more detail: The advanced and retarded solutions of the wave equation are given by the well known advanced and retarded potentials given by, in the usual notation, the well known expression
$$A^\mu_{ret(adv)} (x) = \frac{1}{c} \int \frac{j^\mu (x')}{|r - r'|} \delta \left(| r - r'| \mp c(t - t')\right) d^4 x'$$
(The retarded part of which leads to the Lienard Wiechart potential of earlier theory).\\
It can be seen in the above that we have the situation described within the Compton wavelength, wherein there are two equivalent descriptions of the same event--a photon leaving the charge $A$ and reaching the charge $B$ or the photon leaving the charge $B$ and reaching the charge $A$. The above expression for the advanced and retarded potentials immediately leads to the advanced and retarded fields  (\ref{e3}) and (\ref{e9}) of the F-W description except that we now have a rationale for this formulation in terms of the photon mass and the photon compton wavelength rather than the perfect absorber ad hoc prescription. In fact there is now an immediate explanation for this of the Instantaneous Action At a Distance Theory alluded to. In this case the usual causal electromagnetic field would be given by half the sum of the advanced and retarded fields. We note that as the photon mass is so small, the usual theory is still a good approximation.\\
To sum up, the Feynman Wheeler Perfect Absorber Theory required that every charge should interact instantaneously with every other charge in the universe, that is that the universe must be a perfect absorber of all electromagnetic fields emanating from within. If this condition were satisfied, then the net response of all charged particles along the future light cone of the given charge is expressed by an integral that converges. The present paper argues that this ad hoc prescription of Feynman and Wheeler as embodied by the inclusion of the advanced potential is automatically satisfied if we consider the photon to have a small mass $10^{-65}gms$ as deduced by the author elsewhere, and which is consistent with the latest experimental limits-this leading to the effect mentioned by Weinberg within the Compton wavelength, which is really the inclusion of the advanced field as well. In any case the Machian character is evident in this formulation.
\section{Gravitation}
Gravitation in a sense is a form of weak electromagnetism. A question that has perplexed us for over a century is, why is gravitation so much weaker than electromagnetism - to the extent given by a factor of $10^{-40}$, in fact. One way in which this can be understood is by realizing that the universe is by and large electrically neutral, because the atoms consist of an equal number of positive and negative charges. Strictly speaking atoms are therefore electrical dipoles.\\
With this background let us consider the following simple model of an electrically neutral atom which nevertheless has a dipole effect. In fact as is well known from elementary electrostatics the potential energy at a distance $r$ due to the dipole is given by
\begin{equation}
\phi = \frac{\mu}{r^2}\label{exa}
\end{equation}
where $\mu = eL, L \sim 10^{-8}cm \sim 10^3l \equiv \omega l, e$ being the electric charge of the electron for simplicity and $l$ being the electron Compton wavelength. (There is a factor $cos \, \Theta$ with $\mu$, but on an integration over all directions, this becomes an irrelevant constant factor $4\pi$.)\\
Due to (\ref{exa}), the potential energy of a proton $p$ (which approximates an atom in terms of mass) at the distance $r$ (much greater than $L$) is given by
\begin{equation}
\frac{e^2 L}{r^2}\label{exb}
\end{equation}
As there are $N \sim 10^{80}$ atoms in the universe, the net potential energy of a proton due to all the dipoles is given by
\begin{equation}
\frac{Ne^2 L}{r^2}\label{exc}
\end{equation}
In (\ref{exc}) we use the fact that the predominant effect comes from the distant atoms which are at a distance $\sim r$, the radius of the universe.\\
We now use the well known Eddington formula,
\begin{equation}
r \sim \sqrt{N}l\label{exd}
\end{equation}
If we introduce (\ref{exd}) in (\ref{exc}) we get, as the energy $E$ of the proton under consideration
\begin{equation}
E = \frac{\sqrt{N}e^2\omega}{r}\label{exe}
\end{equation}
Let us now consider the gravitational potential energy $E'$ of the proton $p$ due to all the other $N$ atoms in the universe. This is given by
\begin{equation}
E' = \frac{GMm}{r}\label{exf}
\end{equation}
where $m$ is the proton mass and $M$ is the mass of the universe.\\
Comparing (\ref{exe}) and (\ref{exf}), not only is $E$ equal to $E'$, but remembering that $M = Nm$, we get back in this fine tuned model, famous electromagnetism-gravitation ratio,
$$\frac{e^2}{Gm^2} = \sqrt{N}$$
Here again, the Machian role of all the $N$ particles of the universe, comes into play.\\
It may be mentioned that Einstein himself was much influenced by Mach's ideas in his formulation of the General Theory, which however did not in any way validate it \cite{penrose}.

\end{document}